\documentclass[prx,twocolumn,superscriptaddress]{revtex4-2}

\usepackage{amsmath}
\usepackage{amssymb}
\usepackage{amsthm}
\usepackage{amsfonts}
\usepackage{listings}
\usepackage{physics}
\usepackage{enumerate}
\usepackage{latexsym}
\usepackage{color}
\usepackage{bm}
\usepackage{hyperref}
\hypersetup{
 pdfnewwindow=true, colorlinks=true,
 linkcolor=blue, anchorcolor=blue,
 citecolor=blue, filecolor=blue,
 menucolor=blue, urlcolor=blue}

\usepackage{psfrag}

\usepackage{bm}
\usepackage{graphicx}
\usepackage{subfigure}

\usepackage{multirow}
\usepackage{array}

\RequirePackage[normalem]{ulem} 
\RequirePackage{color}\definecolor{RED}{rgb}{1,0,0}\definecolor{BLUE}{rgb}{0,0,1}

\newcommand{\bk}{{\bf k}}

\def\eg{{\it e.g.}\ }

\input{epsf}

\newcommand{\nc}{\newcommand}
\nc{\webirvsp}{\href{https://github.com/zjwang11/irvsp}{\texttt{IRVSP}} }
\nc{\webirtb}{\href{https://github.com/zjwang11/irvsp}{\texttt{ir2tb}} }
\nc{\webirpw}{\href{https://github.com/zjwang11/ir2pw}{\texttt{ir2ph}} }
\nc{\webchecktopmat}{\href{https://www.cryst.ehu.es/cryst/checktopologicalmagmat}{\texttt{Check Topological Mat}}}
\nc{\webposabr}{\href{https://github.com/zjwang11/UnconvMat/blob/master/src_pos2aBR.tar.gz}{\texttt{POS2ABR}} }
\nc{\webUnconvMat}{\href{http://tm.iphy.ac.cn/UnconvMat.html}{\texttt{UnconvMat}} }
\nc{\online}{\href{http://tm.iphy.ac.cn/UnconvMat.html}{online}}

\begin{document}


\tolerance 10000

\newcommand{\vk}{{\bf k}}

\draft

\title{Three-dimensional topological ferroelectrics}

\date{\today} 


\author{Haohao Sheng}
\affiliation{Beijing National Laboratory for Condensed Matter Physics,
and Institute of Physics, Chinese Academy of Sciences, Beijing 100190, China}
\affiliation{University of Chinese Academy of Sciences, Beijing 100049, China}

\author{Sheng Zhang}
\affiliation{Beijing National Laboratory for Condensed Matter Physics,
and Institute of Physics, Chinese Academy of Sciences, Beijing 100190, China}
\affiliation{University of Chinese Academy of Sciences, Beijing 100049, China}

\author{Zhong Fang}
\affiliation{Beijing National Laboratory for Condensed Matter Physics,
and Institute of Physics, Chinese Academy of Sciences, Beijing 100190, China}

\author{Hongming~Weng}
\affiliation{Beijing National Laboratory for Condensed Matter Physics,
and Institute of Physics, Chinese Academy of Sciences, Beijing 100190, China}
\affiliation{Condensed Matter Physics Data Center, Chinese Academy of Sciences, Beijing 100190, China}

\author{Zhijun~Wang}
\email{wzj@iphy.ac.cn}
\affiliation{Beijing National Laboratory for Condensed Matter Physics,
and Institute of Physics, Chinese Academy of Sciences, Beijing 100190, China}
\affiliation{Condensed Matter Physics Data Center, Chinese Academy of Sciences, Beijing 100190, China}

\begin{abstract}
Three-dimensional (3D) topological ferroelectric (FE) insulators, in which topological and FE orders naturally coexist, enable field-controlled spintronic devices. In this work, we predict a new structure of bismuth monohalides Bi$_4X_4$ ($X$ = Br, I), denoted $\gamma$ phase, and demonstrate that it is an ideal 3D topological FE insulator. Systematic first-principles calculations confirm the stability and synthesizability of $\gamma$-Bi$_4X_4$. Although the noncentrosymmetric $\gamma$ phase crystallizes in the space group $Cmc2_1$ with no symmetry-based classifications/indicators, the nontrivial topology can be characterized by the spin Chern number (SCN). Spin-resolved Wilson loops show the $s_z$ SCN $C_{s_z}=2$, indicating the spin-resolved topology of a 3D quantum spin Hall insulator state. 
The $z$-direction polarization can be switched by interlayer sliding, requiring only crossing a small energy barrier. Finally, we design an electrically controlled spin-filter device on bilayer films that can generate a switchable spin-polarized current. Combining a single-phase crystal, a sizable band gap, and robust band topology against FE switching, these bismuth monohalides serve as a prototype of intrinsic 3D topological FE insulators, providing an ideal platform for realizing new nonvolatile functionalities in spintronic devices.

\end{abstract}

\maketitle

\paragraph*{Introduction.} 
Three-dimensional (3D) topological insulators~\cite{FK-PhysRevLett.98.106803,TI-RevModPhys.82.3045,TI-RevModPhys.83.1057,QSH-HgTe-doi:10.1126/science.1148047,TI-Bi2Se3-nphys1270} and topological crystalline insulators~\cite{TCI-PhysRevLett.106.106802,TCI-SnTe-ncomms1969,TCI-SnTe-nmat3828} exhibit emergent gapless surface or edge states protected by the bulk band topology, offering a material platform with unprecedented electronic properties. 
Owing to spin–momentum locking and suppression of backscattering, these boundary modes enable robust, low-dissipation charge and spin transport that has attracted intense interest for spintronic applications~\cite{TI-spin-PhysRevLett.119.077702,TI-spin-nature13534,TI-spin-nphys3833}.
When such nontrivial topological states coexist with ferroelectric (FE) order, the resulting phase is referred to as a topological FE insulator.
These metallic boundary modes can create an intrinsic short-circuit condition that strongly prevents the depolarization field.
Most interestingly, the topological boundary modes become controllable by an external electric field.

To date, research on topological FE insulators remains at an early stage, and several important challenges persist.
First of all, many efforts focus on two-dimensional (2D) heterostructures with FE switching of topological states, such as In$_2$S$_3$/In$_2$Se$_3$~\cite{In2Se3-In2S2-D2MH00334A}, Sb/In$_2$Se$_3$~\cite{FEtp-Sb-In2Se3-acs.nanolett.0c04531}, and MnBi$_2$Te$_4$/In$_2$Te$_3$~\cite{FEtp-MnBiTe-In2Te3-s41535-025-00800-4}. However, these systems possess very small band gaps (below 0.04 eV), making them susceptible to thermally activated carriers and electrical breakdown. Moreover, the experimental realization of such heterostructures with precisely engineered and lattice-matched interfaces remains a significant challenge~\cite{hete-j.isci.2022.103942,hete-10.1038/nature12385}.
Next, the nontrivial topological state of several topological FE candidates does not survive across the entire FE switching pathway, \eg $\alpha$-Bi monolayer~\cite{Bi-alpha-DQSHI-PhysRevB.105.195142,Bi-alpha-FE-nature}. 
A gapless phase transition point appears along the FE switching pathway, allowing unwanted metallic states to emerge and thereby undermining reproducible, nonvolatile FE devices.
Lastly, 3D FEs are invulnerable to the depolarization field~\cite{depolar-PhysRevB.8.5126,depolar-nature01501}.
Only a few fine-tuned 3D materials are predicted theoretically, such as strained CsPbI~\cite{CsPbI-acs.nanolett.5b04545}, antiferroelectric KMnBi family~\cite{KMBi-AFEtp-PhysRevLett.119.036802,KMgBi-tp-PhysRevLett.117.076401}, and SnTe during its FE phase transition process~\cite{SnTe-FEtp-PhysRevB.90.161108}. 
The intrinsic candidate is lacking.
In order to utilize topological FEs for designing nonvolatile devices, it is essential to discover intrinsic 3D topological FE insulators with sizable band gaps and robust band topology.

Sliding ferroelectricity has been introduced~\cite{slideFE-acsnano.7b02756,BN.Science1,slideFE-PhysRevLett.130.146801} and sliding-induced topological FE metals have been reported in the $T_d$ phase van der Waals (vdW) layered compounds MoTe$_2$ and WTe$_2$~\cite{1T1-MoTe2-wzj-PhysRevLett.123.186401,Td-MoTe2-wzj-PhysRevLett.117.056805,1T1-Td-MoTe2-Huang2019-nc,Td-WTe2-nature15768}. 
FE switching in their bilayers has been observed experimentally~\cite{2layerFE-MoTe2-10.1038,2layerFEexp-WTe2.Nature}. 
In addition, the family of vdW layered bismuth monohalides Bi$_4X_4$ ($X$ = Br, I) has attracted attention.
Experimentally synthesized members include low-temperature $\alpha$-Bi$_4$I$_4$~\cite{alpha-beta-BiI-first-zaac.19784380104,alpha-beta-BiI-natures41586-019-0927-7}, high-temperature $\beta$-Bi$_4$I$_4$~\cite{alpha-beta-BiI-first-zaac.19784380104,alpha-beta-BiI-natures41586-019-0927-7}, and $\alpha$-Bi$_4$Br$_4$ (denoted $\alpha'$ phase for clarity) \cite{1T1-BiBr-first-zaac.19784380105,1T1-BiBr-STM-BrxIx-ncs41467-025-56593-4}.
The structure of $\alpha'$ phase is different from that of $\alpha$-Bi$_4$I$_4$, which is clearly illustrated in Ref.~\cite{PtTe2-shh-PhysRevB.108.104109}. 
Their topological properties have been extensively studied~\cite{BiBr-1layer.nl,alpha-beta-BiI-PhysRevX.11.031042,1T1-BiBr-NMs41563-022-01304-3,alpha-BiI-djz-PhysRevX.14.041048}.
However, all these Bi$_4X_4$ phases are centrosymmetric and consequently lack FE polarization.

In this work, we predict a new $\gamma$ phase of Bi$_4X_4$, and demonstrate that it is an ideal 3D topological FE insulator.
First, first-principles calculations confirm that $\gamma$-Bi$_4X_4$ is stable and synthesizable, and further indicate that $\gamma$-Bi$_4$Br$_4$ is the ground state.
Second, the $s_z$ spin Chern number (SCN) is calculated to be $C_{s_z}=2$, indicating the spin-resolved topology of a 3D quantum spin Hall insulator (QSHI) state. 
Third, $\gamma$-Bi$_4X_4$ is found to be a vdW layered FE with $z$-direction polarization that can be switched by interlayer sliding.
The natural coexistence of topological and FE orders in the single crystal $\gamma$-Bi$_4X_4$, together with a sizable band gap and robust topology, enables field-controlled spintronic devices.
Finally, we propose an electrically controlled spin-filter device on bilayer films that can generate a switchable spin-polarized current.

\begin{figure}[!t]
\centering
\includegraphics[width=0.8\linewidth]{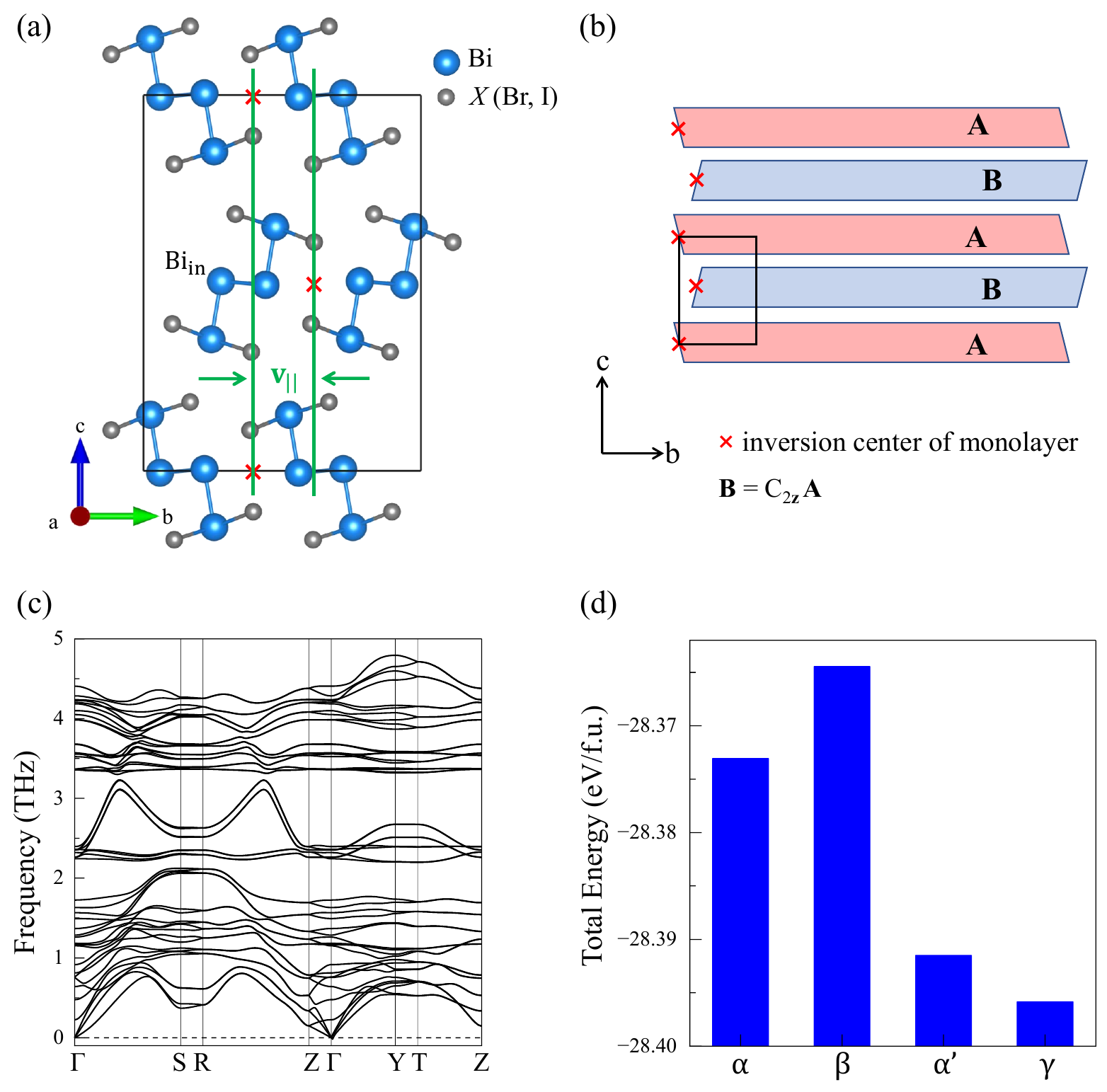}
\caption{(Color online) 
Crystal structure and stability of $\gamma$-Bi$_4$Br$_4$.
(a) Crystal structure. $a$, $b$, and $c$ represent the conventional cell vectors. The red crossings  ``x'' indicate the inversion center of each layer. 
(b) Schematic diagram of vdW stacking. 
(c) Phonon spectrum.
(d) The total energies per formula unit for different phases.
} \label{fig-stru}
\end{figure}

\paragraph*{Crystal structure and stability.} 
The new phase $\gamma$-Bi$_4X_4$ is shown in Fig.~\ref{fig-stru}(a). It is formed by a vdW stacking of the Bi$_4X_4$ monolayers (lying in the $xy$ plane) along the $z$ direction.
As illustrated in Figs.~\ref{fig-stru}(a, b), in this stacking, two adjacent layers are related by a $C_{2z}$ symmetry, forming an AB stacking ($\pi$-bilayer) arrangement.
As reported in Ref.~\cite{PtTe2-shh-PhysRevB.108.104109}, the centrosymmetric monolayers can generate the FE $\pi$-bilayer via interlayer sliding. The similar paradigm also applies to the 3D vdW layered bulk crystals~\cite{1T1-MoTe2-wzj-PhysRevLett.123.186401}.
The sliding vector $\textbf{v}_{||}$ is defined as the in-plane difference between the inversion centers of the adjacent layers. 
After relaxation, $\textbf{v}_{||} = (0.00, 0.22)$ (with respect to the lattice vectors of the conventional cell) is obtained for $\gamma$-Bi$_4$Br$_4$ in Fig.~\ref{fig-stru}(a). Details of the calculation methods are provided in Section \textcolor{blue}{A} of the Supplementary Material (SM).
This vdW layered phase breaks inversion symmetry and belongs to a nonsymmorphic space group $Cmc2_1$, which contains mirror $M_x$, screw rotation $ \widetilde{C}_{2z} \equiv \left\{ C_{2z} \,\middle|\, 0, 0, 0.5 \right\}$, and glide mirror $ \widetilde{M}_{y} \equiv \left\{ M_{y} \,\middle|\, 0, 0, 0.5 \right\}$ symmetries. 
It is worth noting that the $\gamma$ phase is an orthorhombic structure, with two layers in a conventional cell. Therefore, every other layer is aligned well in the $z$ direction [Fig.~\ref{fig-stru}(b)], which is quite different from the reported experimentally synthesized Bi$_4$Br$_4$ ($\alpha'$ phase; Section \textcolor{blue}{B} of the SM).
Thus, we suggest that this structural distinction can be observed in scanning transmission electron microscopy (STEM).

Phonon spectrum calculations show no imaginary modes in Fig.~\ref{fig-stru}(c), confirming that $\gamma$-Bi$_4$Br$_4$ is dynamically stable.  
\emph{Ab initio} molecular dynamics simulations further confirm thermal stability.
Additionally, its formation energy above the convex hull is 0.033 eV/atom, indicating that $\gamma$-Bi$_4$Br$_4$ is likely experimentally synthesizable (Section \textcolor{blue}{C} of the SM).
As shown in Fig.~\ref{fig-stru}(d), $\gamma$-Bi$_4$Br$_4$ is the ground state among all known Bi$_4X_4$ configurations.
The results of Bi$_4$I$_4$ show that the $\alpha$ phase has the lowest energy among them (Section \textcolor{blue}{D} of the SM), which is consistent with its low-temperature phase.
However, the computed phonon spectrum shows that $\gamma$-Bi$_4$I$_4$ is also dynamically stable.
As the Bi$_4X_4$-family compounds of the $\gamma$ phase share the same properties, we take Bi$_4$Br$_4$ as an example in the main text. 

\begin{figure}[!t]
\centering
\includegraphics[width=1.1\linewidth]{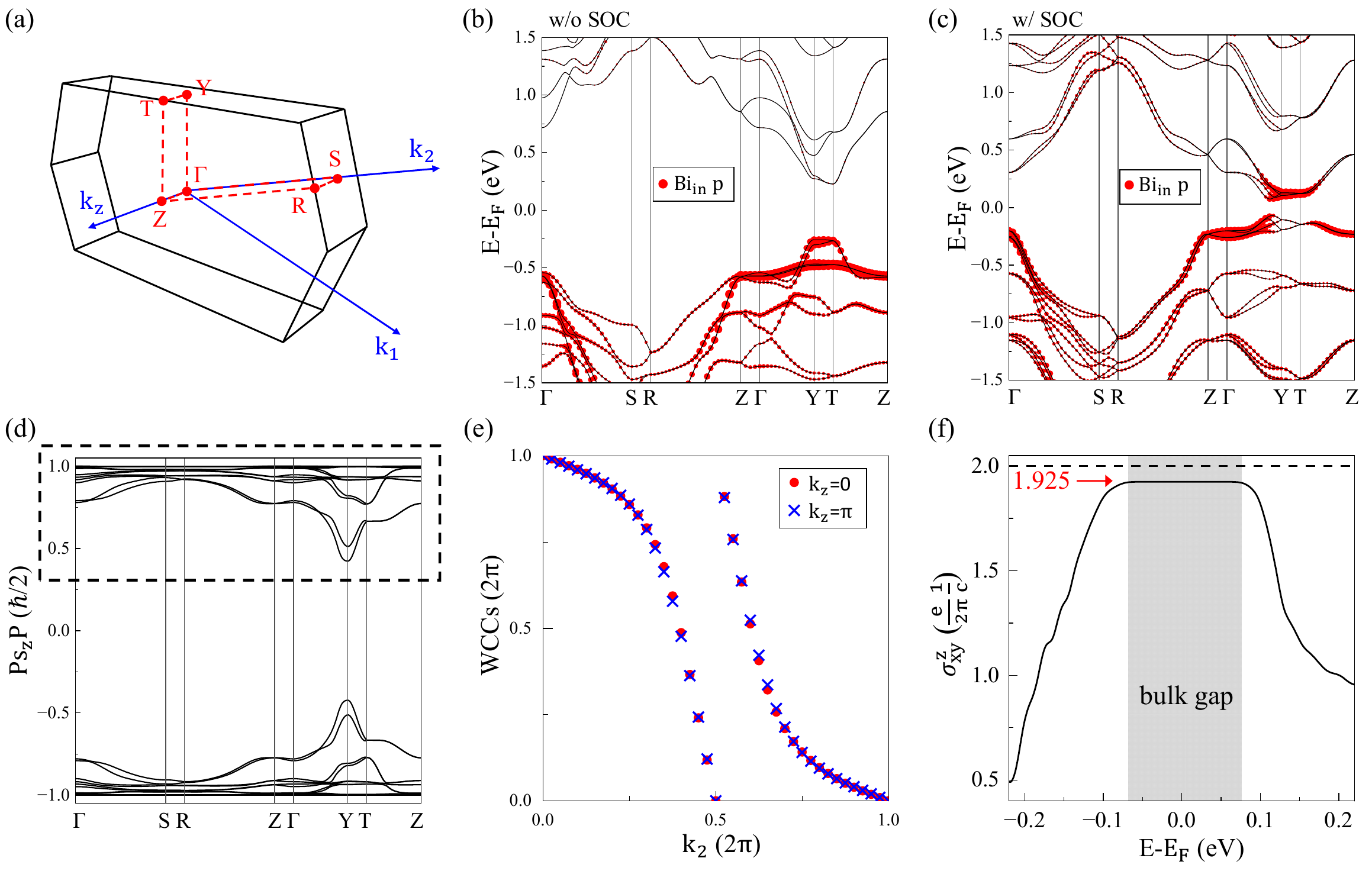}
\caption{(Color online) 
Electronic structure, spin spectrum, and 3D QSHI state of $\gamma$-Bi$_4$Br$_4$.
(a) Brillouin Zone and high symmetry $k$-points of primitive cell.
(b,c) Electronic band structure (b) without (w/o) SOC and (c) with (w/) SOC. The size of red dots represents the weight of the Bi$_{in}$ $p$ orbitals.
Here, Bi$_{in}$ are the middle Bi atoms that form a zigzag chain in each layer [Fig.~\ref{fig-stru}(a)].
(d) Spin spectrum for the eigenvalues of the projected spin matrix $Ps_zP$. The black dashed box presents the upper spin bands.
(e) The $k_{1}$-directed spin-resolved Wilson loops of the upper spin bands for the $k_{z}=0$ (red dots) and $k_{z}=\pi$ (blue crossing points) planes. The resulting spin-resolved Wannier charge centers (WCCs) after summing all Wannier bands are shown.
(f) Intrinsic bulk spin Hall conductivity $\sigma_{xy}^{z}$ as a function of chemical potential. $c$ is the lattice constant in the $z$ direction.
} \label{fig-tp}
\end{figure}

\paragraph*{Band structure and topology.}
From the Perdew-Burke-Ernzerhof band structure in Fig.~\ref{fig-tp}(b), Bi$_4$Br$_4$ exhibits an indirect band gap of $E_g\sim 0.5~eV$. Upon considering spin-orbit coupling (SOC), the gap is significantly reduced to $E_g\sim 0.15~eV$ in Fig.~\ref{fig-tp}(c).
From the weight of the Bi$_{in}$ $p$ orbitals, we can clearly see that the SOC induces the band inversion of two pairs of time reversal (TR)-related bands on the whole $\rm Y$-$\rm T$ line, indicating the potential nontrivial topology.
As SG $Cmc2_1$ has no symmetry-based classifications, the nontrivial topology of Bi$_4$Br$_4$ cannot be simply indicated by any symmetry eigenvalues. 

Then, we find that the band topology is characterized by the $s_z$ SCN.
We have performed the spin-resolved Wilson loop calculations to compute the SCN in a spinful system~\cite{shengli-NJP, djz-spin-PhysRevB.105.224103,Prodan-PRB,spintp-nc}. First, using the $s_z$ spin operator, one can separate the occupied states into two spin subspaces. The projected spin operator is given by $Ps_zP \equiv \frac{\hbar }{2}P(\bk)\hat{\sigma}_zP(\bk)$, where $P(\bk)$ is the projector onto the occupied energy bands. The eigenvalues of the projected spin matrix are presented in Fig.~\ref{fig-tp}(d), which are well separated into the positive part (upper spin bands) and the negative part (lower spin bands). 
Then, we calculate the one-dimensional (1D) Wilson loops for the upper spin bands [black dashed box in Fig.~\ref{fig-tp}(d)] and obtain the spin-resolved Wannier charge centers (WCCs; phases of the eigenvalues of Wilson loop matrices).
The evolution of the WCCs shows that the partial Chern number of the upper spin bands is $C_{s_z}^+ = 2$ for the $k_{z}=0$ plane (red dots) and the $k_{z}=\pi$ plane (blue crossing points) in Fig.~\ref{fig-tp}(e).
The TR symmetry enforces $C_{s_z}^- = -2$ of the lower spin bands.
We find that the SCN is $C_{s_z} = (C_{s_z}^+-C_{s_z}^-)/2 = 2$ for all $k_{z}$ planes, indicating the spin-resolved topology of a 3D QSHI state.

The SCN $C_{s_z} = 2$ produces a \emph{topological} contribution to the spin Hall conductivity (SHC) of $2(\frac{e}{2\pi} \frac{1}{c})$ in the 3D bulk, where $c$ is the lattice constant in the $z$ direction~\cite{spintp-nc}.
We calculate the \emph{intrinsic} SHC $\sigma^{z}_{xy}$ using the Kubo formula~\cite{SHC2004-PhysRevLett.92.126603,SHC2004-PhysRevB.69.235206}
\begin{equation}
\label{SHC}
\begin{aligned}
&\sigma^{z}_{xy} =
\frac{\hbar}{2e} \frac{e^2}{\hbar}
\int_{\mathrm{BZ}} \frac{d^{3}k}{(2\pi)^{3}}
\sum_{n}f_{n}(\mathbf{k})\Omega_{n,xy}^{z}(\mathbf{k}), \\
&\Omega_{n,xy}^{z}(\mathbf{k})=
2 \hbar \sum_{m\neq n}
\frac{-2\mathrm{Im}[
\langle u_{n\mathbf{k}} \vert \hat{J}_{x}^{z} \vert u_{m\mathbf{k}} \rangle
\,
\langle u_{m\mathbf{k}} \vert \hat{v}_{y} \vert u_{n\mathbf{k}} \rangle]}
{
\left( \varepsilon_{n\mathbf{k}} - \varepsilon_{m\mathbf{k}} \right)^{2}
},
\end{aligned}
\end{equation}
where $\hat{J}_{x}^{z} = \tfrac{1}{2}\{\hat{s}_{z}, \hat{v}_{x}\}$ is the spin current operator,  $\hat{s}_{z} = \frac{\hbar}{2}\hat{\sigma}_{z}$ is the spin operator, $\hat{v}_{x} = \frac{1}{\hbar}\frac{\partial H(\mathbf{k})}{\partial k_{x}}$ is the velocity operator, 
$n$ and $m$ are band indices running over all valence and conduction bands,
$\lvert u_{n\mathbf{k}} \rangle$ and $\varepsilon_{n\mathbf{k}}$ are the eigenstates and eigenvalues, $f_{n}(\mathbf{k})$ is the Fermi distribution function, and $\mathrm{BZ}$ denotes the first Brillouin zone. As shown in Fig.~\ref{fig-tp}(f), $\sigma^{z}_{xy}$ exhibits a platform of $1.925(\frac{e}{2\pi} \frac{1}{c})$ within the band gap. 
Our results demonstrate that Bi$_4$Br$_4$ is a 3D QSHI state with a nearly quantized SHC, which is calling for experimental verification.

\begin{figure}[!t]
\centering
\includegraphics[width=0.8\linewidth]{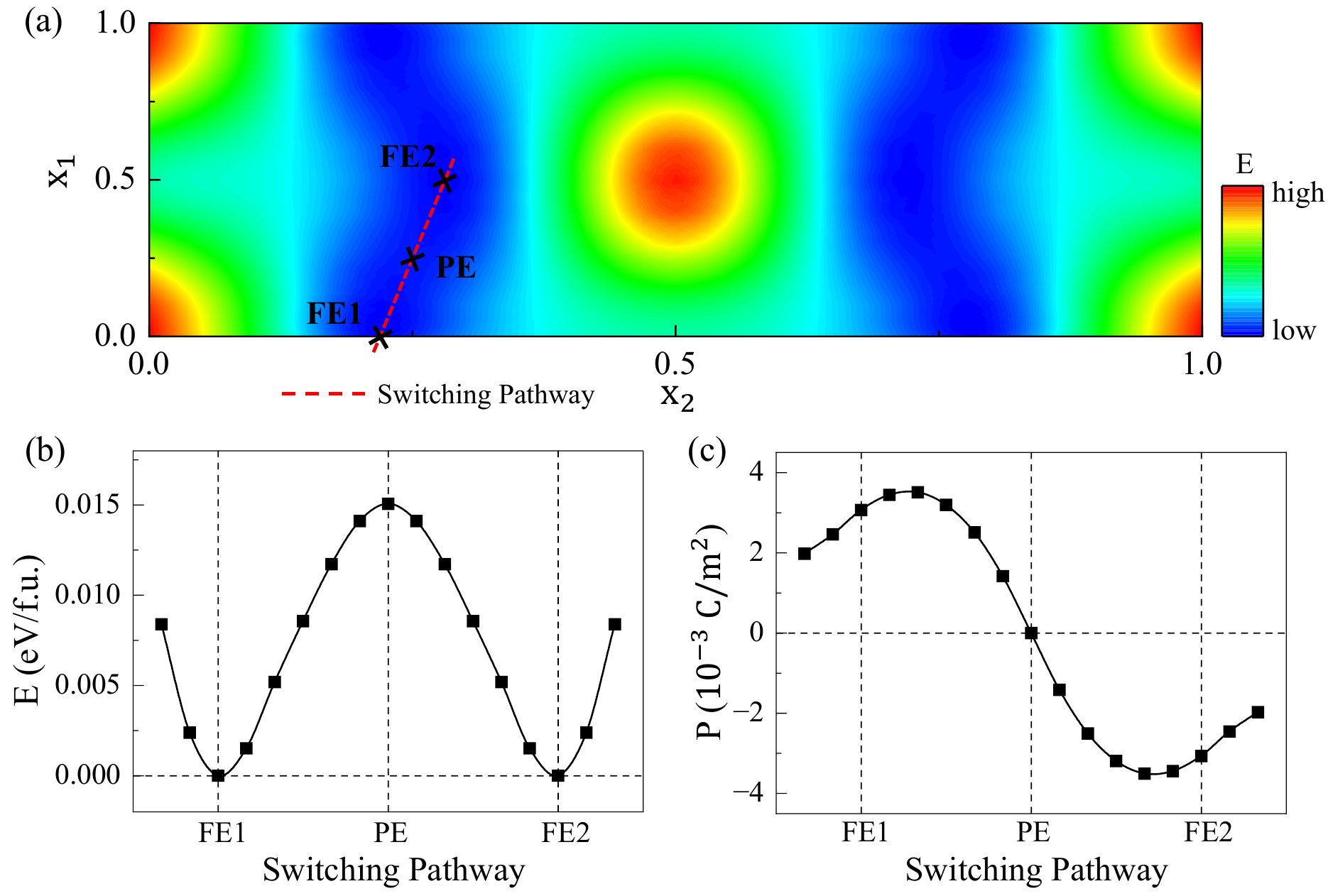}
\caption{(Color online) 
Ferroelectric polarization and switching of $\gamma$-Bi$_4$Br$_4$.
(a) Contour plot of total energy per formula unit as a function of the $\textbf{v}_{||}=(x_1, x_2)$. The FE switching pathway is marked by a red dotted line. 
(b) Energy profile and (c) polarization transition along the pathway. The total energies of two FE phases (FE1 and FE2) are set to zero. PE is the middle paraelectric phase.
} \label{fig-FEbulk}
\end{figure}

\paragraph*{Bulk polarization and ferroelectric.}
Polarization is allowed in Bi$_4$Br$_4$ due to the absence of inversion symmetry.
The $z$-direction (out-of-plane) polarization is calculated to be 3.07$\times$10$^{-3}$ $C/m^2$ in the 3D bulk (2.89$\times$10$^{-3}$ $C/m^2$ for Bi$_4$I$_4$), which is comparable to the reported values of typical vdW stacked $T_d$-phase FEs (such as 2.4$\times$10$^{-3}$ $C/m^2$ for ZrI$_2$~\cite{ZrI2-s41524-021-00648-9}).
To explore FE bistability and switching pathways, we calculate the total energy as a function of $\textbf{v}_{||}$, as shown in Fig.~\ref{fig-FEbulk}(a).
Another FE phase (FE2) with the opposite $z$-direction polarization is obtained at $\textbf{v}_{||}=$(0.50, 0.28), which is linked to FE1 [$\textbf{v}_{||}=$(0.00, 0.22)] by inversion [the base-centered structure yields a translation symmetry of (0.5, 0.5, 0)].
As depicted by the red-colored dashed line, the FE switching can be realized by interlayer sliding.
The energy profile and the polarization transition on the pathway are plotted in Figs.~\ref{fig-FEbulk}(b) and \ref{fig-FEbulk}(c), respectively.
We can clearly see a typical energy double well structure with a low energy barrier of 15 meV per formula unit (meV/f.u.), substantially lower than typical perovskite FEs (such as 170 meV/f.u. for BaTiO$_3$ and 335 meV/f.u. for PbTiO$_3$~\cite{BaTiO-RabeAhnTriscone2007}).
The saddle point [$\textbf{v}_{||}=$(0.25, 0.25)] on the energy surface corresponds to the paraelectric phase (PE) that hosts inversion symmetry. 
Moreover, the spin-resolved topology of the 3D QSHI state survives across the entire FE switching pathway, indicating the robust topology.

\begin{figure}[!ht]
\centering
\includegraphics[width=0.8\linewidth]{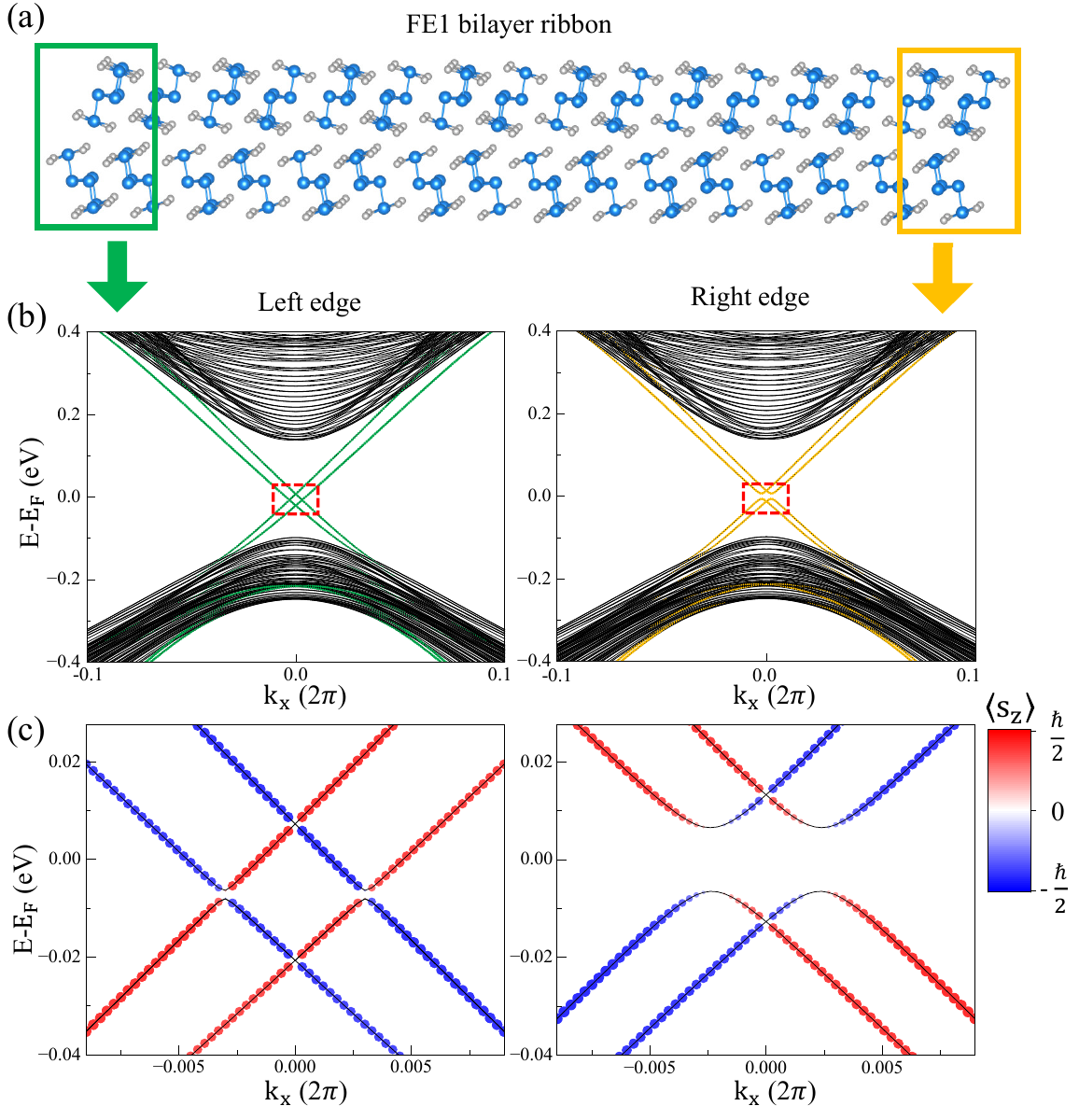}
\caption{(Color online) 
Topological edge states in Bi$_4$Br$_4$ $\pi$-bilayer.
(a) Crystal structure of the FE1 ribbon.
(b) Band structures with SOC of the FE1 ribbon. The left edge states (green dots) and the right edge states (orange dots) are shown in the left panel and the right panel, respectively. 
(c) The zoom-in band structures corresponding to the red dashed boxes in (b). The color scale and the size of dots represent the expectation value of the $s_z$ spin.
} \label{fig-edge}
\end{figure}

\paragraph*{Electrically controlled spin-filter in bilayers.} 
To utilize the unique properties of the topological FE insulator, we design an electrically controlled spin-filter device on bilayer films.
Due to the vdW coupling between layers, the $\pi$-bilayers can be obtained by exfoliation or molecular beam epitaxy. 
The 2D FE metals have been reported in the $\pi$-bilayers of $1T$-PtTe$_2$ and $1T'$-MoTe$_2$ families~\cite{PtTe2-shh-PhysRevB.108.104109,PtTe2-exp-CPL070717}.
Here, the Bi$_4$Br$_4$ $\pi$-bilayer is a 2D FE insulator~\cite{PtTe2-shh-PhysRevB.108.104109}, consisting of two QSHI layers with interlayer sliding. 
We systematically investigate the topological edge states of Bi$_4$Br$_4$ $\pi$-bilayers.
The crystal and band structures of the FE1 ribbon are presented in Fig.~\ref{fig-edge}. Two pairs of edge states are clearly visible within the bulk band gap at each edge, with spin orientations predominantly aligned along the $z$ axis. The interlayer hybridization of these spin-momentum-locked edge states plays a crucial role in determining the spin‑current characteristics of the $\pi$-bilayer device.
On the left edge [left panels of Figs.~\ref{fig-edge}(b, c)], two pairs of edge states exhibit a \emph{negligible} hybridization gap (less than 1 meV), yielding two forward-propagating spin-up channels and two backward-propagating spin-down channels. In contrast, the right-edge states exhibit a \emph{sizable} hybridization gap [15 meV; right panels of Figs.~\ref{fig-edge}(b, c)].
These features indicate that the FE1 $\pi$-bilayer functions as a spin-filter at low temperature~\cite{device-jianghua-PhysRevB.99.195422}: when an unpolarized current is injected into the ribbon, a purely spin-up-polarized current is generated along a single left edge.

\begin{figure}[!t]
\centering
\includegraphics[width=0.9\linewidth]{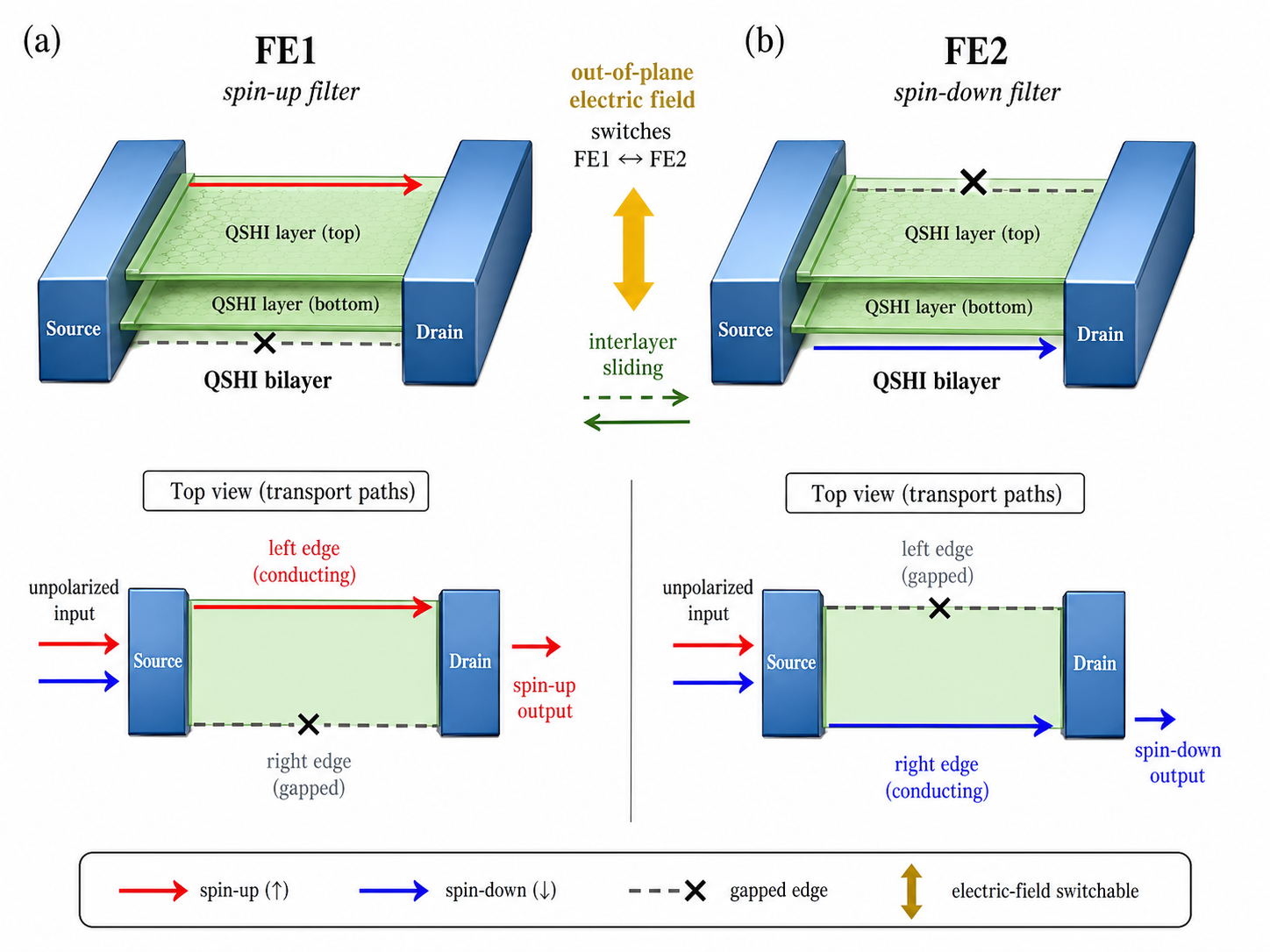}
\caption{(Color online) 
Electrically controlled spin-filter device. (a, b) A fully spin-polarized current is realized, where opposite spin polarizations are spatially locked to the different conducting edges and can be switched by an out-of-plane electric field.
} \label{fig-device}
\end{figure}

The schematic of the FE-controlled spin filter device is shown in Fig.~\ref{fig-device}(a).
After applying an $z$-direction electric field, the bilayer can be switched from the FE1 phase to the FE2 phase via interlayer sliding. 
The FE2 ribbon can show a spin‑filter behavior with opposite polarization: an unpolarized current now conducts solely along the right edge and becomes purely spin‑down-polarized, as shown in Fig.~\ref{fig-device}(b).
Thus, these results demonstrate an electrically controlled spin-filter device that generates a switchable spin‑polarized current, where the opposite spin polarizations are spatially locked to the different conducting edges and can be switched electrically.

\vspace{20em}
\paragraph*{Conclusion and discussion.} 
In summary, we predict a new $\gamma$ phase of bismuth monohalides Bi$_4X_4$ and demonstrate that it is an ideal 3D topological FE insulator with a sizable band gap and robust band topology. 
Due to the unique vdW stacking and the sliding-induced inversion-symmetry breaking, we propose that the novel $\gamma$ phase can be identified experimentally by the observation of the alignment between every other layer in STEM and by the observation of a nonzero second harmonic generation (SHG) signal. 
The spin-resolved topology of the 3D QSHI state is indicated by the nonzero SCN $C_{s_z}=2$. 
The computed bulk SHC $\sigma_{xy}^{z}=1.925(\frac{e}{2\pi} \frac{1}{c})$ suggests that the nearly quantized 3D QSH effect can be expected in experiment.
Moreover, the sliding ferroelectricity is found in the topological $\gamma$-Bi$_4X_4$ compounds. We suggest that the FE switching can be observed experimentally in the flakes.
Finally, we propose an electrically controlled spin-filter device on the bilayer films capable of producing a switchable spin-polarized current. 
Overall, the family of $\gamma$-Bi$_4X_4$ can serve as a very promising material platform for the electrically controlled spin-filter device and other electrically controlled, nonvolatile quantum spintronic devices.


\ \\
\paragraph*{Acknowledgments.}
This work was supported by the National Natural Science Foundation of China (Grant No. 12188101), the National Key R\&D Program of China (Grant No. 2022YFA1403800),  and the Center for Materials Genome.


%

\clearpage

\end{document}